\documentclass[a4paper,11pt,twocolumn]{article}

\usepackage{graphicx}
\usepackage{url}
\usepackage{ulem}
\usepackage{rsc} 
\renewcommand{\emph}{\textit}

\date{}

\title{Unveiling the atomic structure of single-wall boron nanotubes}

\author{
Jens Kunstmann,\thanks{
Dr.~Jens Kunstmann, Department of Chemistry, Columbia University, 3000 Broadway New York, NY 10027, USA; corresponding e-mail: jk3610@columbia.edu 
\newline
Dr.~Jens Kunstmann, Dr.~Viktor Bezugly, Hauke Rabbel, Prof.~Dr.~Gianaurelio Cuniberti, Institute for Materials Science and Max Bergmann Center of Biomaterials, TU Dresden, 01062 Dresden, Germany
\newline
Dr.~Viktor Bezugly, Prof.~Dr.~Gianaurelio Cuniberti, Center for Advancing Electronics Dresden, TU Dresden, 01062 Dresden, Germany
\newline
Prof.~Dr.~Mark H.~R\"ummeli, Leibniz Institute for Solid State and Materials Research Dresden (IFW), Helmholtzstr. 20, 01069 Dresden, Germany and Institut f\"ur Festk\"orperphysik, TU Dresden, 01062 Dresden, Germany
} \
Viktor Bezugly, Hauke Rabbel,\\
Mark H. R\"ummeli, Gianaurelio Cuniberti
}

\begin{document}


\maketitle

\begin{abstract}
Despite recent successes in the synthesis of boron nanotubes (BNTs), the atomic arrangement of their walls has not yet been determined and many questions about their basic properties do remain.
Here, we unveil the dynamical stability of BNTs by means of first-principles molecular dynamics simulations. We find that free-standing, single-wall BNTs with diameters larger than 0.6 nm are thermally stable at the experimentally reported synthesis temperature of 870$^\circ$C and higher.
The walls of thermally stable BNTs are found to have {a variety of different} mixed triangular-hexagonal morphologies. 
{Our results substantiate the importance of mixed triangular-hexagonal morphologies as a structural paradigm for atomically thin boron.}
\end{abstract}

\section{Introduction}

Low-dimensional structures of current interest such as monolayer transition metal chalcogenides, graphene, carbon nanotubes or silicon nanowires are very promising systems for the possible realization of future nanotechnologies. After the prediction of stable, quasi-planar and tubular clusters of elemental boron \cite{Boustani1997c,Boustani1997,Quandt2005,Boustani2011a}, which could later be confirmed experimentally \cite{Zhai2003,Kiran2005,Oger2007}, materials scientist have started to search for boron nanostructures similar to graphene and carbon nanotubes.
Several models for boron sheets and nanotubes (BNTs) with different underlying lattice structures have been proposed \cite{Boustani1999,Kunstmann2006,Lau2006a,Tang2007,Singh2008,Yang2008} and first successes in growing pure BNTs were reported \cite{Ciuparu2004,Liu2010,Liu2011}. 
In contrast to carbon nanotubes, which  can be either semi-conducting or metallic depending on their diameter and chiralities, BNTs are predicted to be metallic only \cite{Boustani1999,Lau2006a,Kunstmann2006} and highly conductive. \cite{Bezugly2011,Bezugly2013}
For small-diameter BNTs related to the $\alpha$-sheet (diameter $<$ 1.7 nm), some density functional theory (DFT) calculations predict that the nanotubes are semiconducting due to a curvature-induced out-of-plane buckling of certain atoms.\cite{Singh2008,Yang2008,Tang2010}
However, recent calculations at higher levels of theory (second order M\o{}ller-Plesset perturbation theory) and with van der Waals corrected DFT  show that the buckling might be an artefact of standard DFT \cite{Szwacki2010,Gunasinghe2011}. Without buckling all BNTs are indeed metallic.

This feature could make BNTs excellent candidates for future nanometer-scale conducting elements.
Recently Liu \textit{et al.} reported conductivity measurements on large-diameter (10 to 40 nm) multi-walled BNTs which seem to confirm this prediction.\cite{Liu2010} 
Besides the metallic sheets and nanotubes there is a growing body of literature on semi-conducting boron nanostructures that are probably related to the known bulk crystal structures \cite{Cao2001,Wu2001,Otten2002,Zhang2002,Tian2010}. 
These developments indicate the rise of a very promising branch of nanoscience based on boron nanostructures \cite{Tian2010,Quandt2008a}.

\begin{figure}[t]
\includegraphics[width=\columnwidth]{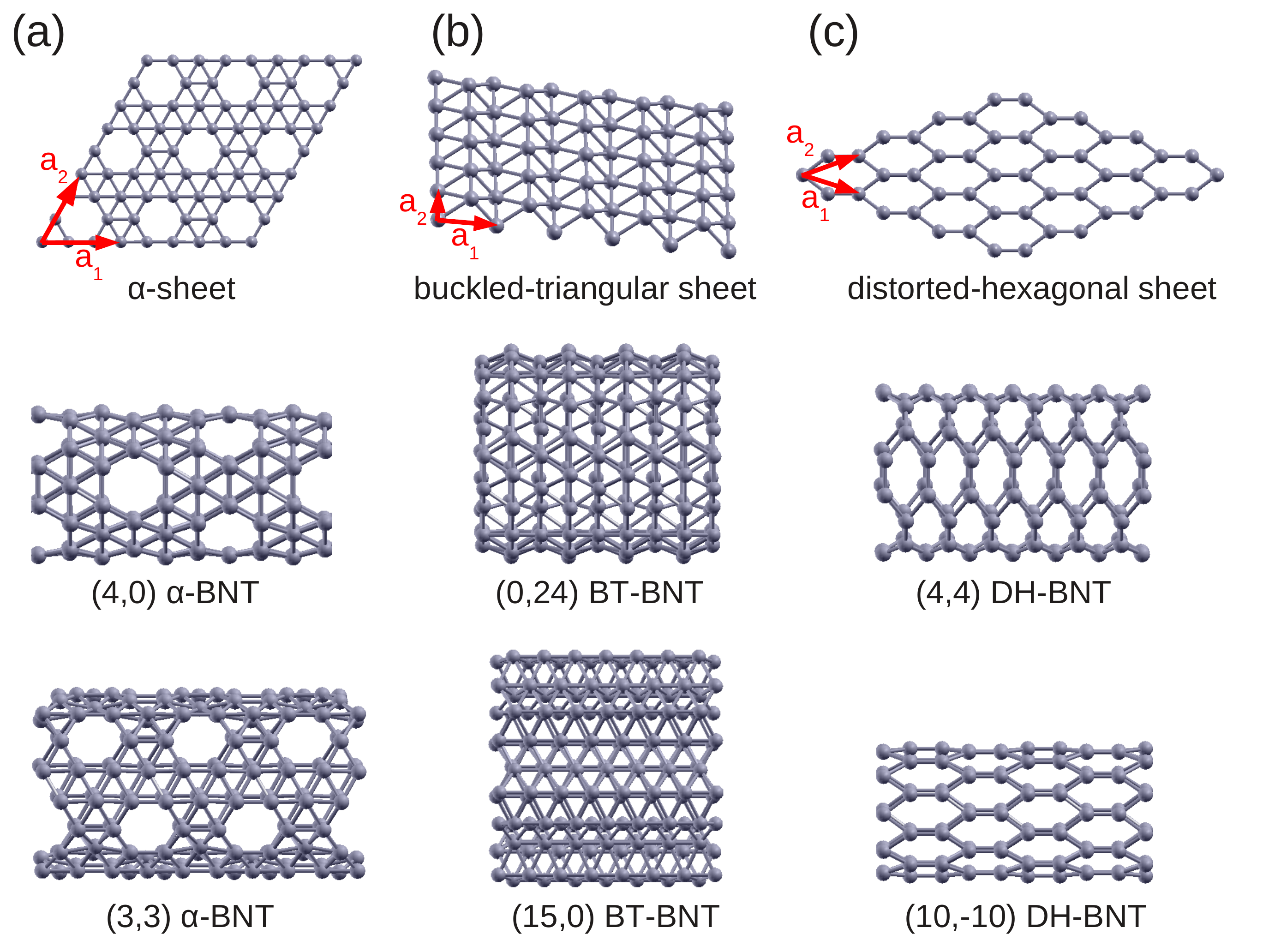}
\caption{\label{fig:structures}
Possible atomic structures of boron sheets and the related boron nanotubes of armchair (second line) and zigzag (third line) type. (a) Structures derived from the $\alpha$-sheet, (b) the buckled triangular (BT) sheet, and (c) the distorted hexagonal (DH) sheet. We consider the three different structure models (a-c) as the atomic structure of boron nanotubes is not definitely known. {The primitive lattice translations $a_1$ and $a_2$ of the unit cells of the boron sheets are shown in red and the numbers $(n,m)$ are the chiral indices of the nanotubes.}
}
\end{figure}

Despite these early successes in the synthesis and characterization of BNTs, many questions on their structure and physical properties remain open and even their existence is still debated. 
In contrast to carbon or boron-nitride, boron does not form layered bulk structures and therefore the atomic structure of the walls of BNTs still needs to be determined experimentally.
{
The bulk phases of boron are based on three--dimensional frameworks of slightly distorted B$_{12}$ icosahedra (icosahedral boron crystals). Four elemental modifications are confirmed to exist, i.e.,  $\alpha$-rhombohedral, $\beta$-rhombohedral, $\beta$-tetragonal, $\gamma$-orthorhombic \cite{Oganov2009a} boron and a fifths phase, $\alpha$-tetragonal boron \cite{Ekimov2011}, is currently discussed again after it was discarded in the 1970s.
The icosahedral structural unit is also central for boron-hydrogen compounds and it is therefore a generally accepted structural paradigm in boron chemistry.
Deviating from this paradigm it was found that small boron clusters prefer quasi-planar and tubular morphologies \cite{Boustani1997c,Boustani1997,Zhai2003,Kiran2005,Quandt2005,Oger2007,Boustani2011a}. Extending these findings into the nano-domain boron sheets, boron nanotubes \cite{Boustani1999,Kunstmann2006,Lau2006a,Tang2007,Singh2008,Yang2008} and boron fullerenes \cite{Boustani1997a,Szwacki2007} were proposed. However, the existence of non-icosahedral structures, that go beyond isolated, finite-size clusters, is not generally accepted.
}

Nevertheless several models for boron sheets and the related BNTs are currently discussed in the literature (see Figure \ref{fig:structures}). In this article we consider the buckled triangular (BT) sheet,\cite{Kunstmann2006} the distorted hexagonal (DH) sheet\cite{Lau2006a} and the so-called $\alpha$-sheet \cite{Tang2007}. Each sheet represents a favorable representative of three general structural classes: triangular, hexagonal (honeycomb lattice), and mixed triangular-hexagonal (MTH) structures, respectively. As elemental boron exhibits a pronounced polymorphism each of these sheets could in principle be the precursor of BNTs. Furthermore it is important to note that all related structures (sheets and BNTs) have metallic properties.

The first indication that the theoretically predicted structures might indeed have been synthesized came from the calculation of the work function that agrees with the experimental value for large-diameter BNTs to high accuracy only for $\alpha$-sheet but not for the DH and BT sheets \cite{Bezugly2011}. 

Planar structures of the MTH class were introduced by Tang \textit{et al.} \cite{Tang2007} and are characterized by their hexagonal hole density. That is the number of hexagonal holes $N_\mathrm{holes}$ divided by the number of atoms in the related hole-free triangular sheet, or:
\begin{equation}
 \eta = \frac{N_\mathrm{holes}}{N_\mathrm{atoms} + N_\mathrm{holes}},
\label{eqn:eta}
\end{equation} 
where $N_\mathrm{atoms}$ is the number of atoms in the actual sheet with holes. $\eta$ ranges from $0$ to $1/3$ and the two limiting cases correspond to the triangular and hexagonal sheets, respectively. The highest cohesive energy among all boron sheets was initially obtained by density functional theory (DFT) within the local density approximation (LDA) for the $\alpha$-sheet with $\eta = 1/9$ \cite{Tang2007}. 
The stability of the $\alpha$-sheet can be understood by a self-doping picture, i.e., while a flat triangular boron sheet is electron-rich and a hexagonal boron sheet is electron-deficient, a regular combination of triangular and hexagonal structural elements, as in the $\alpha$-sheet, creates an optimal balance and thus a stable structure \cite{Tang2007,Tang2009}.
However the energy minimum is relatively shallow and 
a multitude of different sheets with $\eta \approx 1/9$ and different spatial arrangements of the holes are discussed in the literature. All of these  sheets have very similar cohesive energies and therefore the results for the absolute energy minimum (and the precise value of $\eta$) vary slightly depending on the actual electronic structure method and DFT exchange-correlation functional \cite{Wu2012,Penev2012,Yu2012}. The variety of nearly isoenergetic structures is again a realization of boron's well known polymorphism. 
Therefore it is likely that a multitude of MTH structural patterns with $\eta \approx 1/9$ will coexist and disorder might be an important factor, yet to be explored in the context of boron sheets and BNTs.
For B$_{80}$ boron fullerenes \cite{Szwacki2007}, that are structurally related to the $\alpha$-sheet, the influence of disorder was already considered. Pochet \textit{et al.} found that disordered MTH arrangements on the surface of the fullerene can have lower total energies than ordered arrangements \cite{Pochet2011} and they also speculate about the stability of disordered BNTs. The same authors also find the energy landscape of boron clusters to be glasslike which seems to explain the experimental difficulties in the synthesis of boron nanostructures \cite{De2011}. 
{
Furthermore they find that B$_{80}$ or B$_{100}$ boron clusters do not prefer hollow, fullerene-like structures but rather disordered cages with an icosahedron inside. These results seem to confirm the structural paradigm of the icosahedron and speak against the stability of MTH structures.
}

In order to characterize disordered and defective sheets and nanotubes the parameter $\eta$ is too limited, because its definition is based on triangular structures and hexagonal holes, only.
We therefore propose to characterize MTH structures with non-hexagonal holes
by their atomic surface density, i.e.,
\begin{equation}
 n_\mathrm{2D} = \frac{N_\mathrm{atoms}}{S},
\label{eqn:n}
\end{equation} 
where $S$ is the surface area of a sheet or a nanotube wall and $N_\mathrm{atoms}$ is the number of atoms within that surface. In principle $\eta$ and $n_\mathrm{2D}$ can be converted by
$n_\mathrm{2D} = c_{\eta} (1-\eta)$. The constant $c_{\eta}$ is chosen to reproduce an $\eta$ value of a specific structure, for instance $\eta=1/9$ for the $\alpha$-sheet corresponds to $n_\mathrm{2D} = 0.372$ {\AA$^{-2}$} with $c_{\eta} = 0.419$. 
However $n_\mathrm{2D}$ and $\eta$ are conceptually different quantities that cannot readily be converted. $\eta$ characterizes merely the morphology of a 2D surface but characteristic bond lengths of a certain structure are not taken into account. The surface density $n_\mathrm{2D}$ on the other hand depends on both bond lengths and the surface morphology. 
{However, structures with the same morphology can have different surface densities, e.g., comparing a flat and a buckled surface. Furthermore, we point out that the surface area is not uniquely defined on the atomic scale. Here we determine the surface area by Delaunay triangulation, i.e., $S$ is the sum of all triangular areas drawn between atom triples in a way that no holes and overlapping regions exist.}
Throughout this paper we will use the surface density $n_\mathrm{2D}$ to characterize boron sheets and BNTs. While $\eta$ ranges between 0 (flat triangular) and $1/3$ (flat hexagonal lattice), the corresponding limiting surface densities (obtained with DFT/PBE, see methods section) are 0.396 {\AA$^{-2}$} and 0.272 {\AA$^{-2}$}, respectively.

\begin{figure}[t]
\includegraphics[width=\columnwidth]{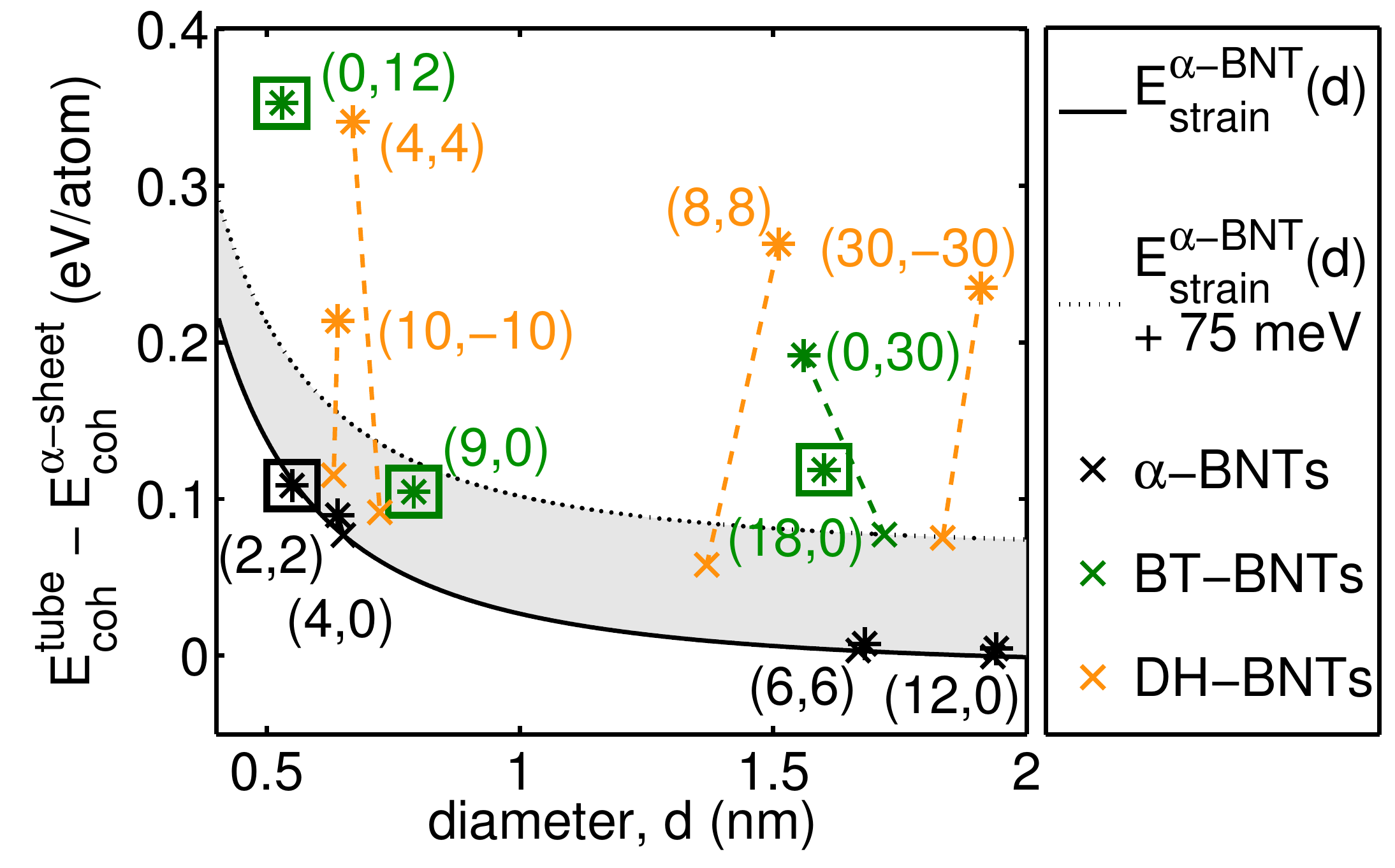}
\caption{\label{fig:energies} 
Stability, structural transitions and collapse of boron nanotubes (BNTs) at high temperatures: the graph shows cohesive energies $E_\mathrm{coh}^\mathrm{tube}$ of different BNTs relative to the cohesive energy $E_\mathrm{coh}^\mathrm{\alpha-sheet}$ of the $\alpha$-sheet before ($\ast$) and after ($\times$) \textit{ab initio} molecular dynamics simulation at 870$^\circ$C. We consider $\alpha$-BNTs, buckled-triangular (BT) BNTs and distorted-hexagonal (DH) BNTs. During the simulations  the systems are either stable (nearby symbols), they transform from BT or DH initial structures into mixed triangular-hexagonal BNT structures (symbols connected by broken lines) or they collapse (symbols with boxes). 
Final structures are found to be within an energy window of +75 meV/atom relative to the energy of perfect $\alpha$-BNTs (indicated by their diameter-dependent strain energy curve $E_\mathrm{strain}^\mathrm{\alpha-BNT}(d)$).
}
\end{figure}

So far, theoretical investigations on boron sheets and BNTs have mostly focused on determining the zero temperature ground state structure by structural searches and optimizations \cite{Wu2012,Penev2012,Yu2012,Liu2013}. 
However to prove the existence and the thermal stability of atomic structures it is necessary to consider their dynamical properties, i.e., lattice vibrations. Lau \textit{et al.}~and Wu \textit{et al.}~showed by DFT-based phonon calculation that several simple sheet models are dynamical unstable because their phonon dispersions exhibit imaginary frequencies \cite{Lau2008a,Wu2012}, low energy MTH structures however were shown to be dynamically stable \cite{Wu2012}. 
As phonons are collective harmonic vibrations about the structure's equilibrium positions, high-temperature properties and phase transitions cannot easily be accessed. 
In order to judge whether BNTs derived from the MTH sheets are suitable candidates to explain the experimental observations of BNTs \cite{Ciuparu2004,Liu2010,Liu2011}, it is necessary to prove that they are thermally stable  at the synthesis temperature of 870$^\circ$C. 
{
The glasslike energy landscape of boron clusters and the strong tendency of boron to form 3D icosahedral structures \cite{De2011} suggests that under strong thermal motion initial nanotubues would lose their tubular shape and transform into disordered structures with some icosahedra inside.
Boron nanotubes can only exist if this does not happen. 
Therefore this article is trying to answer the following questions: 
1. Can boron nanotubes with atomically thin walls and non-icosahedral geometries exist?
2. If yes, what is their atomic structure?	
}

As result we show by molecular dynamics (MD) simulations based on DFT that free-standing, single-walled BNTs with MTH structures, diameters larger than ca.~0.6 nm 
are indeed thermally stable at the synthesis temperature. During the course of the MD simulation BNTs related to the $\alpha$-sheet are fully stable and have a mean surface density of $n_\mathrm{2D} = 0.357$ {\AA$^{-2}$}.
Initial DH and BT structures are not stable and either collapse or transform into partially defective, MTH structures with broken crystallinity but local atomic order and a mean surface density of $n_\mathrm{2D} = 0.340$  {\AA$^{-2}$}. 
The observation of such remarkable structural transitions shows that atomically thin boron tends to form MTH structures when it is subject to external (synthesis) conditions that force it into a tubular shape. 
Thus MTH {morphologies are indeed} a paradigm in boron chemistry for atomically thin structures which amend the paradigm of the B$_{12}$ icosahedron that is considered for bulk structures and boranes. Furthermore our results indicate that disorder and defects might play a much more important role in boron nanostructures  than in carbon or boron-nitride based nanostructures.

\section{Results and Discussion}

\begin{figure}[t]
\includegraphics[width=.9\columnwidth]{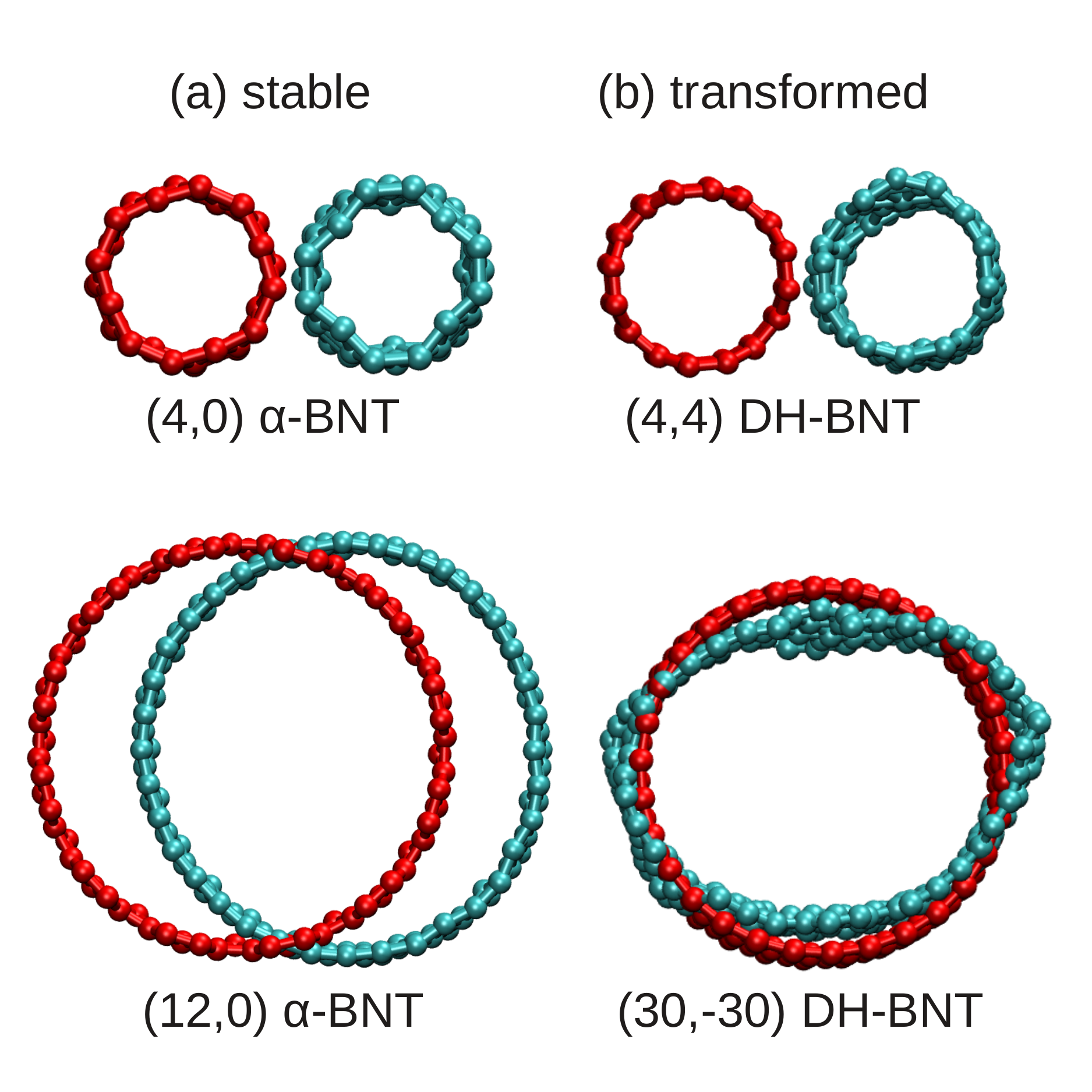}
\caption{\label{fig:topview} 
(a) Stable and (b) structurally transformed 
boron nanotubes with different diameters seen along their main axis.
Red represents ideal, initial structures before the MD simulation and green are the final, annealed and optimized structures after the MD simulation.
}
\end{figure}

\begin{figure}[t]
\includegraphics[width=\columnwidth]{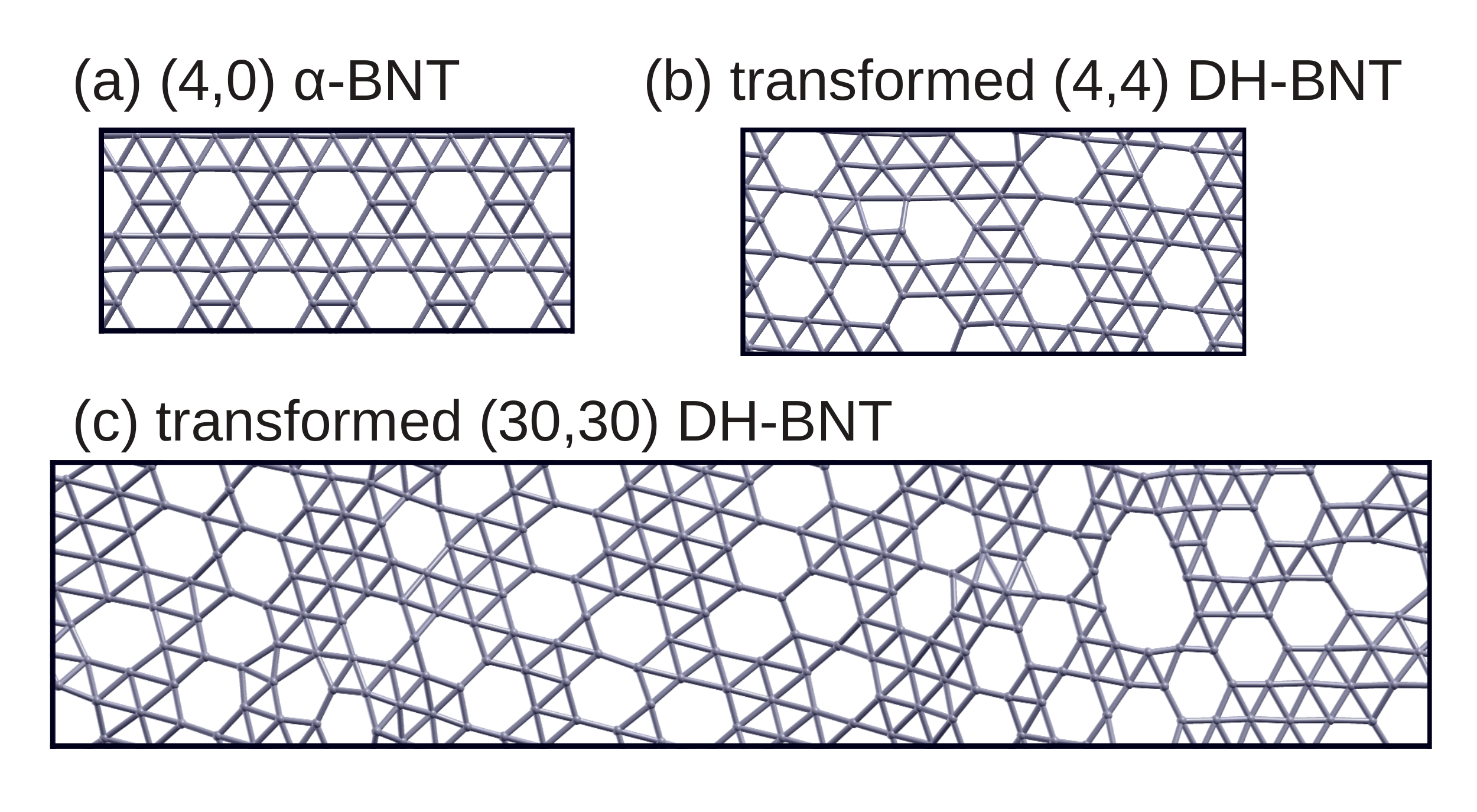}
\caption{\label{fig:unrolled}
The atomic structure of the walls of some considered boron nanotubes, unrolled into a plane. The black lines represent the units of repetition (the horizontal direction corresponds to the circumference and the vertical direction to the axial direction of the actual BNT). 
The transformed structures (b) and (c) clearly resemble the ideal $\alpha$-BNT (a), where triangular and empty hexagonal motives prevail. The basic difference to ideal $\alpha$-BNTs is the armorphous structure and the occurrence of non-hexagonal holes.
}
\end{figure}

\begin{figure}[t]
\includegraphics[width=.8\columnwidth,trim=20 20 20 60]{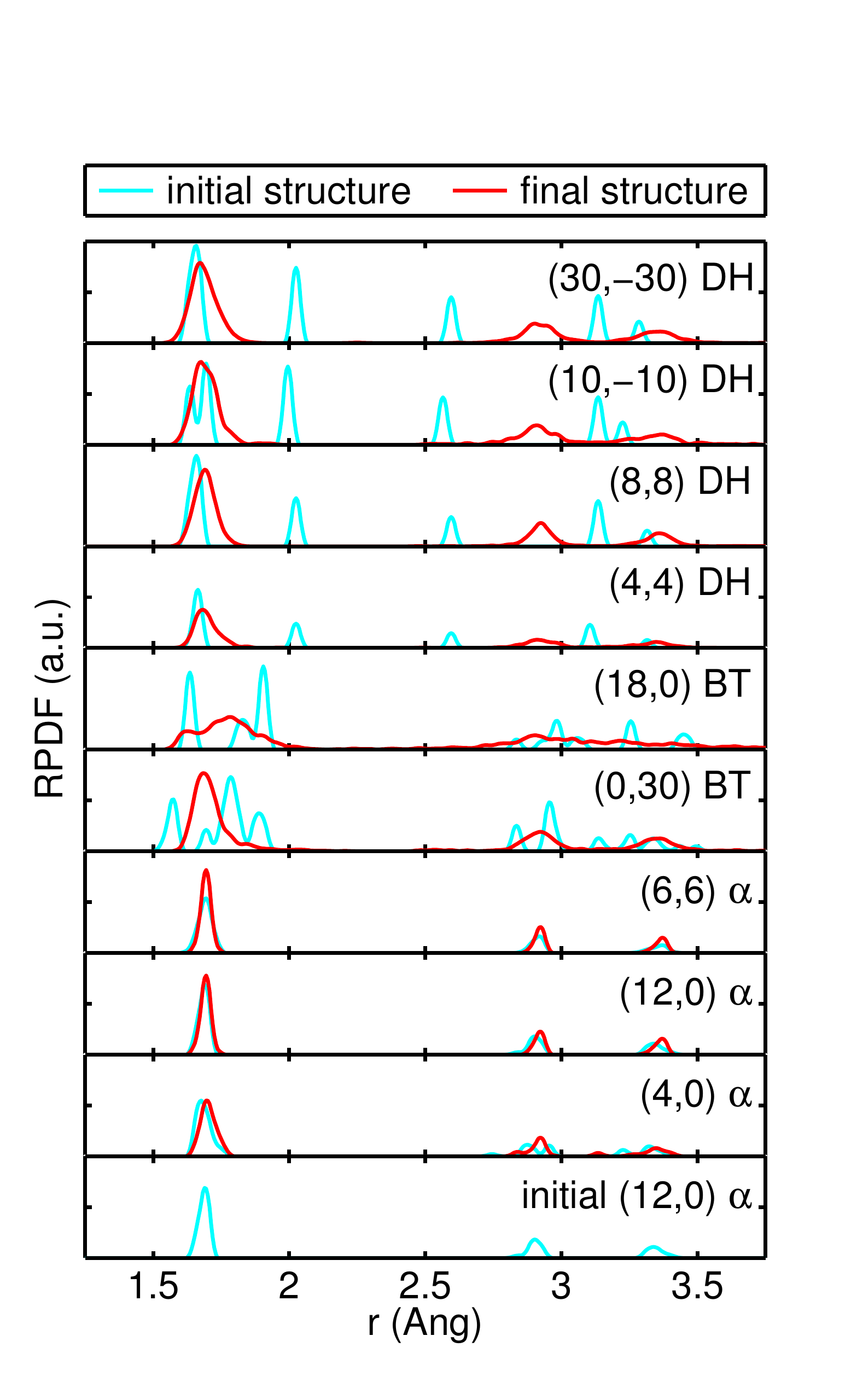}
\caption{\label{fig:rpdf}
Structural transitions of boron nanotubes visualized by the change of their radial pair distribution function (RPDF): before (cyan lines) and after (red lines) the MD simulation. The RPDFs of nanotubes within one of the three structural classes ($\alpha$, BT or DH) have similar characteristic peaks. But the RPDFs of nanotubes from different structural classes are clearly distinguishable. The RPDF of all stable, final structures is similar to one of the prototypical (12,0) $\alpha$-nanotube shown in the bottom panel. This indicates that the structures of all thermally stable nanotubes are relatively similar to $\alpha$-BNTs structures. Mind that the BT-zigzag (18,0) is a collapsed structure.
}
\end{figure}

In order to study the thermal stability of BNTs we consider 4 BNTs of each of the three structural class, i.e., buckled triangular (BT), distorted hexagonal (DH) and $\alpha$-BNTs, summing up to a total of 12 BNTs. The 4 BNTs of each class are an armchair and a zigzag nanotube with a small ($<$ 0.8 nm) and a big ($>$ 1.5 nm) diameter. Armchair and zigzag BNTs of each structural class are illustrated in Figure \ref{fig:structures}. The chiral indices $(n,m)$ of a BNT of each class are defined by the wrapping vector $\mathbf{W} = n \mathbf{a_1} + m \mathbf{a_2}$ that spans the circumference of the nanotube, and $\mathbf{a_1}$ and $\mathbf{a_2}$ refer to the primitive lattice translations of the underlying boron sheet {as indicated in Figure \ref{fig:structures}}.

The dynamics of the system is simulated by \textit{ab initio} MD simulations based on DFT within the Generalized Gradient Approximation for the exchange-correlation potential. For the details on the simulations we refer to the methods section at the end of this article. 
Boron exhibits very complex and quite flexible chemical bonding patterns (mixtures of covalent two-center and multi-center bonds) that are very hard to capture in 
approximate quantum methods. For static ground state properties and harmonic vibrational frequencies some progress within the framework of the density functional tight-binding method has been made recently \cite{Grundkotter-Stock2012}. However, to obtain reliable structural dynamics at high temperatures it is necessary to perform the MD simulations on the computationally expensive \textit{ab initio} level, where the chemical bonding is correctly represented for any given instantaneous geometry. Therewith the high-temperature behavior and structural transitions of the BNTs can be correctly described.

The basic outcome of this study is summarized in Figure \ref{fig:energies} where we plot the cohesive energies (also known as binding energies or atomization energies) $E_\mathrm{coh} = E_\mathrm{tot}/N - E_\mathrm{atom}$ of different BNTs relative to the cohesive energy $E_\mathrm{coh}^\mathrm{\alpha-sheet}$ of the $\alpha$-sheet. $E_\mathrm{tot}$, $N$ and $E_\mathrm{atom}$ are the total energy of a system, the number of atoms in the system and the total energy of a single B atom, respectively (note that with this definition the cohesive energy has a negative sign).
As the cohesive energy of the $\alpha$-sheet is smaller  than the one of the BNTs (indicating more binding energy), the most stable structures are shown in the lower part of the figure and the stability decreases in the vertical direction. 
The difference of the cohesive energies between a nanotube and the corresponding sheet is the so called strain (or curvature) energy $E_\mathrm{strain} = E_\mathrm{coh}^\mathrm{tube} - E_\mathrm{coh}^\mathrm{sheet}$. It describes the amount of energy per atom needed to roll a flat sheet into a specific tube. For sheets with nearly isotropic in-plane properties (like the $\alpha$-sheet) the strain energies obeys $E_\mathrm{strain}(d) = C/d^2$ \cite{Kunstmann2007}, where $d$ is the tube's diameter and $C=0.03694$ eV~nm$^2$/atom is our fit to the final results for $\alpha$-BNTs. The value for $C$ agrees well with previous results \cite{Singh2008,Bezugly2011}. The strain energy curve of $\alpha$-BNTs is shown as black continuous line in the lower part of Figure \ref{fig:energies}.

During the simulations at 870$^\circ$C the different initial structures (indicated by $\ast$ symbols in Figure \ref{fig:energies}) are either stable, structurally transform into MTH-BNTs (symbols connected by broken lines) or collapse (symbols with boxes). Figure \ref{fig:topview} gives examples for initial (red) and final structures (green) illustrating these types of behaviors. The initial and final geometries of the stable structures in (a) are hardly distinguishable, the cross-sections of transformed ones in (b) can be non-circular due to the presence of defects and disorder and the collapsed BNTs lose their tubular shape as will be discussed below.

The most stable structures are the $\alpha$-BNTs. Only the small diameter (2,2) $\alpha$-BNT collapsed and the other $\alpha$-BNTs kept their initial structure at all times. Their dynamical stability, shown by the MD simulations, is further supported by their high cohesive energies, indicating high chemical binding energies. The cohesive energies of $\alpha$-BNTs set the boundary for the most stable MTH structures that are indicated by the strain energy curve $E_\mathrm{strain}(d)$ in Figure \ref{fig:energies}. 
In Figure \ref{fig:unrolled}(a) the structure of the wall of one $\alpha$-BNTs after the MD simulation is shown. It is indeed identical the structure of the $\alpha$-sheet in Figure \ref{fig:structures}(a). Figure \ref{fig:topview}(a) shows examples of stable small and large-diameter $\alpha$-BNTs. Final $\alpha$-BNTs differ from their initial counterparts only by small surface modulations that also give rise to small energy differences in Figure \ref{fig:energies}. For details about these surface modulations we refer the reader to another work \cite{Tang2010}.
While the planar $\alpha$-sheet has a surface density of $n_\mathrm{2D} = 0.372$ {\AA$^{-2}$}, the values of the three stable $\alpha$-BNTs are $n_\mathrm{2D} = 0.357 \pm 0.003$ {\AA$^{-2}$}, with the difference coming from the out-of plane surface modulations that increase the surface area and decrease the density compared to the flat $\alpha$-sheet.

A second class of simulated structures undergo a structural transition. These are all DH-BNTs and the (0,30) BT-BNT. This finding is quite remarkable as the MD simulations provide final structures of MTH type without the need to artificially impose that. This shows that boron tends to form MTH structural patterns if it is subject to boundary conditions that enforce an overall planar or tubular shape. 
MTH structures are a specific realization of structures predicted by the Aufbau principle by I. Boustani, that described the general morphology of small boron clusters \cite{Boustani1997b}. We think that they can be considered as a fundamental paradigm in boron chemistry for atomically thin systems. 

The transformed MTH-BNTs are shown in Figure \ref{fig:topview}(b) and \ref{fig:unrolled}(b),(c). These structures are very similar to the $\alpha$-sheet.
However the hexagons occur in a disordered way, i.e., the crystallinity is broken, but local atomic order, represented by similar triangular and hexagonal structural motives, is maintained.
Furthermore polygonal holes with 4, 5, 7, 8  and even 11 vertices can be found, which we will subsequently refer to as lattice defects. 
As a matter of computational feasibility all our simulations are done using periodic boundary conditions (the unit cells are indicated as black lines in Figure \ref{fig:unrolled}). Therefore the occurrence of MTH-BNTs with broken crystallinity and local atomic order after a structural transition {is most likely} a result of incommensurate simulation cells that do not allow crystalline structures to be formed. 
{The unit cells of most of the transformed BNTs are either incommensurate with the alpha-sheet lattice or would require axial contractions bigger than 10\% that are not accessible by our simulation because the size of the unit cells is fixed. Only the (8,8) DH-BNT could transform into a (5,5) $\alpha$-BNT with -0.6\% axial strain and a 9\% radial contraction. However within the 20 ps simulation time we did not observe this transition.}
The fact that the crystalline $\alpha$-BNTs are generally more stable (in terms of cohesive energies) than the transformed, disordered MTH-BNTs (see Figure \ref{fig:energies}) seem to support {further the influence of the finite size unit cells.} 
Pochet \textit{et al.} pointed out the role of disorder in boron fullerenes and they conjectured that disordered BNTs could be more stable than ordered ones \cite{Pochet2011}. Our results do not seem to confirm this conjecture.
It appears that the complexity of the energy landscapes of the 1D BNTs and the one of 0D boron clusters \cite{De2011} differ.
However tranformed MTH-BNTs are dynamically stable in our MD simulations indicating the possible existence of BNTs surface structures that lack long-range order.
The surface densities of the 5 transformed BNTs is $n_\mathrm{2D} = 0.340 \pm 0.004$ {\AA$^{-2}$}, i.e., less dense than the $\alpha$-BNTs ($n_\mathrm{2D} = 0.357$ {\AA$^{-2}$}). Although the surface structures of the transformed BNTs differ strongly and they are irregular and partially defective, their surface densities are practically the same. This apparently surprising result seem to reflect the tendency of the BNTs to optimize the ratio between electron-rich filled hexagons and electron-deficient empty hexagons (self-doping) during the course of the MD simulation. 
The fact that the surface density of the $\alpha$-BNTs is higher than the one of the transformed ones (see above) might indicate that the self-doping is not fully optimized in $\alpha$-BNTs, that keep their $\alpha$-type geometries during the MD simulation.  Recent works of Penev \textit{et al.}~and Yu \textit{et al.}~for 2D boron sheets showed that DFT calculations based on the PBE exchange-correlation functional find the absolute minimum of the cohesive energy vs. $\eta$ convex hull to be at $\eta=1/8=0.125$ and not at $\eta=1/9=0.11$ ($\alpha$-sheet) \cite{Penev2012,Yu2012}. As a bigger $\eta$ value corresponds to a smaller surface density their results for 2D boron sheets and our results for 1D BNTs seem to indicate a similar trend. 
Interestingly, the cohesive energies of the transformed MTH-BNTs are found to vary over a sizable energy range of 75 meV/atom relative to the energy of perfect $\alpha$-BNTs (see Figure \ref{fig:energies}).

The dynamics of the structural transitions from DH or BT structures to MTH ones initially occurs on very short time scales and then continues slowly as a process that eliminates defects and reduces disorder. For more details on the dynamics we refer to Figure S1 of the Supporting Information.
Even though the structural transitions slowly proceeds further with time the basic nanotubes are perfectly stable at 870$^\circ$C. Thus our results show that partially defective, MTH-BNTs with broken crystallinity and local atomic order are thermally stable. 

The fact that BT-BNTs and DH-BNTs either collapse or undergo structural transformations but are never stable rules out these models as candidates to explain the atomic structure of BNTs.
%
In the collapsed structures atoms on opposite walls of the initial nanotube come close to each other during the MD simulation and eventually form covalent bonds. Depending on the system more and more bonds are formed and the structures loose their tubular shape. In some of the collapsed systems 
fragments of B$_{12}$ icosahedra can be found, indicating boron's tendency towards forming bulk structures. However the surface-to-volume ratio of the collapsed systems is is still too big to fully transform into bulk-like systems.
We wish to point out that the structural collapse of zigzag BT-BNTs was predicted by one of the authors earlier \cite{Kunstmann2006}.
Among the collapsed structures there is the (2,2) $\alpha$-BNT where 1-2 covalent bonds are formed between opposite walls. Such effect is not known to happen in carbon nanotubes. The fact that small diameter BNTs are unstable agrees with the experimental observation that only nanotubes with large diameters are found \cite{Ciuparu2004,Liu2010,Liu2011}. We empirically determine the smallest possible diameter of stable BNTs to be at about 0.6 nm.

To further characterize the BNTs in consideration we calculate the radial pair distribution function (RPDF) that gives the number of atoms in a spherical shell of infinitesimal thickness at a distance $r$ from a reference atom. The RPDF peaks at distances that are characteristic to a certain structure and for distances $r<2.2$ {\AA} it indicates characteristic bond lengths.
Nanotubes within one of the three structural classes ($\alpha$, BT or DH) exhibit similar characteristic peaks but the RPDFs of nanotubes from different structural classes are clearly distinguishable. This indicates that the three structural classes exhibit distinguishable, characteristic bond lengths. 
While the initial DH-BNTs and BT-BNTs have several peaks the $\alpha$-BNTs show only one narrow peak at 1.68 {\AA}.
The RPDFs of the transformed MTH nanotubes exhibit broader, single peaks at similar interatomic distances, indicating that all thermally stable MTH nanotubes have similar structures than  $\alpha$-BNTs, which is also obvious from inspecting Figure \ref{fig:unrolled}. 
The relatively narrow bond length distribution for defect-free $\alpha$-structures is rather unusual for boron-rich systems, where the presence of distinct two-center and multi-center bonds usually implies a relatively broad bond length distribution.
In Figure \ref{fig:rpdf} we also show the RPDF of the collapsed (18,0) BT-BNT. As its structure is more irregular the RPDF is characterized by a broader distribution of bond length.

Finally, to test the dynamical stablility of BNTs some of these systems were also simulated at a higher  temperature of 1500 K. The initial structures were the final structures of the $T= 1143$ K (870$^\circ$C) simulation run, if the latter were stable or have structurally transformed. For collapsed ones, the initial, ideal structures were used. We found that $\alpha$-BNTs and BT-BNTs behave the same way as at $T=1143$ K (see Figure \ref{fig:energies}), whereas small-diameter DH-BNTs [(4,4) and (10,-10)] collapsed and large-diameter [(8,8) and (30,-30)] remain to be stable as MTH-BNTs. 
The small-diameter DH-BNTs collapse the same way as described above, i.e., by virtue of their thermal movement  atoms on opposite sides of the nanotube come close to each other and eventually form a  covalent bond. These results seem to suggest that the stability of small diameter MTH-BNTs varies with temperature, the degree of disorder and the number of defects. The influence of these parameters has to be studied in more detail in the future.

\section{Summary and Conclusion}
To summarize, by means of DFT-based molecular dynamics simulations we were able to show that free-standing, singe-walled BNTs are thermally stable at the experimentally reported synthesis temperature of 870$^\circ$C and higher. The latter were found to have mixed triangular-hexagonal (MTH) morphologies and diameters larger than ca.~0.6 nm.
BNTs related to the $\alpha$-sheet are thermally stable. Distorted-hexagonal (DH) and buckled-triangular (BT) BNTs are thermally unstable and during a MD simulation they either collapse or transform into partially defective, MTH-BNTs with broken crystallinity and local atomic order. 
Thus DH and BT-BNTs can be  ruled out as structural models for BNTs. 
Although originating from non-MTH initial structures the transformed MTH-BNTs all have very similar atomic surface densities of $0.340 \pm 0.004$ {\AA$^{-2}$}. This density could reflect a good self-doping balance between the number of electron-rich filled hexagons and the number of electron-deficient empty hexagons in a system.
The chemical binding energies of the crystalline $\alpha$-BNTs are higher than the ones of the MTH-BNTs with broken crystallinity, indicating a tendency of BNTs to prefer crystalline surface structures.
However the thermal stability of the MTH-BNTs with broken crystallinity and local atomic order
shows that, similar to findings in boron bulk structures, disorder and defects might play a more significant role than they do for carbon or boron-nitride nanotubes.
The structural transformations observed during the MD simulations show that boron tends to form MTH structures when it is subject to external (synthesis) conditions that force it into a planar or tubular shape. We therefore believe that MTH morphologies are a structural paradigm for atomically thin boron-rich systems 
and amend the paradigm of the B$_{12}$ icosahedron that is considered for bulk structures and boranes.

\section{Computational Methods}

We simulate infinite, free-standing nanotubes by placing lattice-periodic nanotubes fragments in hexagonal supercells with the periodic direction going along the $z$ axis and a separation of 10 {\AA} between neighboring nanotubes in the $x$ and $y$ directions. The unit cell sizes along the $z$ direction are system dependent and were chosen to be between 8 and 12 {\AA}.

The MD simulations are performed in the NVT ensemble (constant number of particles, volume and temperature) using the Nos\'{e}-Hoover thermostat \cite{Nose1984} with a mass parameter of $\mu = 0$. A Verlet algorithm with a time step of 1 fs is used to integrate Newton's equation of motion. To avoid the appearance of unphysical vibrational modes all simulations start with a 1 ps warm-up phase, where the temperature is linearly risen from 0 K to either 1143 K (870$^\circ$C) or 1500 K. The warm up phase is followed by 20 ps simulation at constant temperature. Stable structures are then annealed by linearly lowering the temperature down to 0 K within 1 ps. The annealed structures are then structurally optimized (including the lattice parameter in the $z$ direction) until all interatomic forces are below 0.01 eV/{\AA} and, in a final static (singe-point) calculations, the total energy is determined.

All of these calculations are based on the density functional theory \cite{Kohn1965} (DFT) within the framework of the generalized gradient approximation (GGA) using the parametrization by Perdew, Burke and Ernzerhof (PBE) \cite{Perdew1996}. We used the Vienna ab initio simulation package (VASP, version 5.2) \cite{Kresse1996,Kresse1996a} employing the projected augmented wave (PAW) method \cite{Blochl1994a,Kresse1999}.
A plane wave basis set with a kinetic energy cutoff of 240 eV was used for the MD simulations and a  cutoff of 320 eV for structural optimizations and calculations of total energies. In all self-consistent calculations the total energies were converged, such that energetic changes were less than 10 meV for the MD simulations and less than 0.1 meV for the structural optimizations and calculations of total energies. In order to ensure that the wave functions of the MD runs are sufficiently converged, the DFT self-consistency loop was required to do at least 4 iterations at each MD time step.
For the k-point sampling of the Brillouin zone, $\Gamma$-point centered grids were used. The k-space integration was carried out with the method of Methfessel and Paxton in first order \cite{Methfessel1989} and a smearing width of 0.1 eV for all calculations. Optimal k-point meshes are individually converged for each system by reducing the changes in the total energy at least below 3 meV/atom.

The surface area $S$ of all BNTs (see equation \ref{eqn:n}) was determined by Delaunay triangulation and properly taking into account the periodicity of the nanotubes fragments. The RPDFs in Figure \ref{fig:rpdf} were calculated from the nanotube fragments with VMD {(software to visualize and analyze molecular structures)}\cite{Humphrey1996} and a radial resolution of 0.03 {\AA}, mapped on a finer radial grid and then smoothened with a running average window of 0.03 {\AA}.

\section{Supporting Information}
Supporting Information is available, showing the time evolution of the potential energy of stable, collapsing and structurally transforming BNTs.

\section{Acknowledgements}
JK, VB and GC acknowledge financial support from the German Research Foundation (DFG) (project KU 2347/2-2). MHR thanks the German Research Foundation (DFG RU1540/15-2).
This work is partly supported by the German Research Foundation (DFG) within the Cluster of Excellence "Center for Advancing Electronics Dresden" and by the European Union (ERDF) and the Free State of Saxony via TP A2 ("MolFunc"/"MolDiagnosik") of the Cluster of Excellence "European Center for Emerging Materials and Processes Dresden" (ECEMP).
The computations were performed at the Center for Information Services and High Performance Computing (ZIH) of  the TU Dresden. 

\bibliographystyle{rsc} 
\bibliography{references}

\end{document}